\documentclass[12pt,a4paper,final]{iopart}

\usepackage{iopams}
\usepackage{graphicx}
\usepackage{xfrac}

\begin{document}

\title{Growth-induced breaking and unbreaking of ergodicity in fully-connected 
spin systems}

\author{Richard G.~Morris}
\address{Theoretical Physics, Department of Physics, The University of Warwick, 
Coventry, CV4 7AL, UK.}
\ead{r.g.morris@warwick.ac.uk}

\author{Tim Rogers}
\address{Centre for Networks and Collective Behaviour, Department of 
Mathematical Sciences, University of Bath, Bath, BA2 7AY, UK.}
\ead{t.c.rogers@bath.ac.uk}


\begin{abstract}
	Two canonical models of statistical mechanics, the fully-connected voter 
	and Glauber-Ising models, are modified to incorporate growth via the 
	addition or replication of spins.  The resulting behaviour is examined in a 
	regime where the timescale of expansion cannot be separated from that of 
	the internal dynamics.  Depending on the model specification, growth 
	radically alters the long-time dynamical behaviour by breaking or 
	unbreaking ergodicity.
\end{abstract}

\pacs{75.10.-b, 05.40.-a, 02.50.Ey}
\submitto{\JPA}

\maketitle
\section{Introduction}
The dynamical behaviour of growing physical systems often involves an interplay 
between \textit{internal dynamics}, arising from the interaction rules between 
the constituent parts, and the \textit{mechanism of growth}, such as addition, 
adsorption or replication.  Illustrative examples of such systems include the 
formation of stars and galaxies \cite{EKMST90,ABL-ZF94}, the growth of 
cancerous tumours \cite{DHRAW00,NB+07} and the ecology of bacterial outbreaks 
\cite{SRH+10}.  Existing theoretical work on the subject of growth is, however, 
largely separated into two specific sub-areas:  \textit{pure-growth processes}, 
concerned with the time dependence of shape and form, such as directed 
percolation, diffusion limited aggregation, and P\'{o}lya-urn random 
reinforcement models \cite{HJH86,FFTV91,MM+96,P07}; and, \textit{phase-ordering 
kinetics}, focusing on the scaling of internal dynamical behaviour such as 
relaxation or coarsening \cite{HF85,AB94,DIRRAB11}.  Neither of these branches 
fully capture the broad range of dynamical behaviours possible when the 
timescales of growth and internal dynamics cannot be separated. Such systems 
comprise a third important class of models, where growth drives the internal 
dynamics of the system.

In order to better characterise this broad class, we concentrate on 
highly-idealised, fully-connected, spin-\sfrac{1}{2} systems; specifically, the 
voter \cite{TML99} and Glauber-Ising models \cite{Kinetic}.  Aside from the 
fact that these systems are amenable to analytical treatment, the benefit of 
such an approach is to demonstrate the effects of growth on models whose 
behaviour at fixed system size is very well understood.  Indeed, incorporating 
a mechanism for the addition or replication of spins strongly affects the 
long-time behaviour of the magnetization: changing the ergodicity 
characteristics and the statistics of polarization flips.  Under the addition 
of random spins, the voter model becomes ergodic, with the long-time 
distribution depending on the rate of growth. Under spin replication dynamics, 
however, the voter model remains non-ergodic with the statistics of the final 
state again depending on the rate of growth. By contrast, it appears that 
growth of any type is sufficient to break ergodicity in the Glauber-Ising 
model, as the potential wells always become absorbing at long times.

\section{Setup}
To start, consider a generic fully-connected spin system. In the absence of 
growth such systems comprise a fixed number $N$ of spins 
$\sigma_1,\ldots,\sigma_{N} \in \{-1,+1\}$, that interact via probabilistic 
rules for spin `flips'. In a typical protocol, spins are updated one at a time, 
with one unit of time corresponding to $N$ spin updates (one `Monte-Carlo 
sweep'). The probability $P(\{\sigma_i\},t)$ that the system is in a given 
state at time $t$, evolves according to a master equation \cite{NGVK}.  For 
fully-connected systems, spin flip rates depend only on the magnetization $x= 
\sum_i^N \sigma_i/N$, and the description of the system is reduced to a 
stochastic jump process in $x$, with jumps of size $2/N$.  For \textit{growing} 
systems however, $N=N(t)$ is itself an increasing function of time and 
therefore the number of spin updates per unit time also increases.  The 
relevant scaling parameter (which must independent of time) is then the 
\textit{initial} system size $N_0\equiv N(t=0)$, which provides a lower bound 
on $N(t)$.  With this in mind, such growing systems can be formally examined by 
expanding the relevant master equation--- now modified for the effects of 
growth--- in powers of $1/N_0$.  The point being that $O(1)$ contributions in 
the resulting expansion describe deterministic internal dynamics for which 
growth is negligible (\textit{i.e.}, corresponding to the limit $N_0\to\infty$) 
and higher order terms give corrections that characterise the dominant effects 
of growth.

For simplicity, we define our growing systems in continuous time. Spins are 
updated independently, with an average of one update per spin per unit time, in 
correspondence with the usual Monte-Carlo time in non-growing simulations.  The 
shorthand $f_u$ ($f_v$) is used for the probability that an up (down) spin is 
flipped, once chosen. At rate $\lambda$, new spins are picked from a reservoir 
and added to the system.  New spins are up with probability $g_u$, or otherwise 
down with probability $g_v=1-g_u$.  The probabilities $f_{u}$, $f_{v}$, $g_{u}$ 
and $g_{v}$ may depend on the magnetization, while the growth rate $\lambda$ is 
generally a function of the system size.  The regime of interest is when the 
timescales of growth and spin flips cannot be separated: that is, when newly 
added spins interact--- but do not equilibrate--- with existing spins, before 
more are added.  For this reason, $\lambda$ is required to scale like 
$N^\alpha$ with exponent $\alpha\in(-1,1)$.

Before writing down the master equation, it should be remarked that to specify 
the state of a (fully-connected) growing system requires two order parameters, 
rather than one.  The magnetization $x$ and the system size $N$ comprise one 
possible choice; here, the pair $(u,v)$ is used instead, where $uN_0$ and 
$vN_0$ are the total number of up and down spins, respectively.  These 
variables have the advantage that their change is measured in constant steps of 
$1/N_0$, unlike magnetization, whose step-size gets smaller as the system gets 
larger.  The master equation is
\begin{eqnarray}
		\frac{\textrm{d}}{\textrm{d}t}P(u,v,t)=
		\bigg[&N_0\left(\mathcal{E}_{u}^{-1/N_0}\mathcal{E}_{v}^{+1/N_0}-1\right)
		vf_{v}
		+N_0\left(\mathcal{E}_{u}^{+1/N_0}\mathcal{E}_{v}^{-1/N_0}-1\right)
		uf_{u}\nonumber\\
		&+\left(\mathcal{E}_{u}^{-1/N_0}-1\right)\lambda g_u
		+\left(\mathcal{E}_{v}^{-1/N_0}-1\right)\lambda g_v\bigg]P(u,v,t),
	\label{eq:M-E}
\end{eqnarray}
where we have made use of step operators $\mathcal{E}$ with action 
$\mathcal{E}_p^r F(p,q)=F(p+r,q)$, for a generic function $F$. For small values 
of the step increment $r$, this action is well-approximated by Taylor series, 
since $F(p+r,q)\approx F(p,q)+rF_p(p,q)+r^2F_{pp}(p,q)/2$. This allows for the 
expansion of the master equation for large $N_0$ in a standard way \cite{NGVK}.  
Neglecting $O(N_0^{-2})$ terms, a Fokker-Planck equation is obtained from 
(\ref{eq:M-E}) that is equivalent to the following Langevin pair: 
\begin{eqnarray}
		\frac{\textrm{d} u}{\textrm{d} t} &= \left(v f_{v} - u f_{u}\right) + 
		\frac{\lambda}{N_0} g_u + \sqrt{\frac{u f_{u} + v 
		f_{v}}{N_0}}\,\eta(t),\label{eq:Langevin1}\\
		\frac{\textrm{d} v}{\textrm{d} t} &= \left(u f_{u} - v f_{v}\right) + 
		\frac{\lambda}{N_0} g_v - \sqrt{\frac{u f_{u} + v 
		f_{v}}{N_0}}\,\eta(t),
	\label{eq:Langevin2}
\end{eqnarray}
where $\eta(t)$ is a single source of standard Gaussian white noise, understood 
in the It\={o} sense.  Note: whilst Langevin equations are preferred for 
clarity of discussion, we make repeated use of the fact that a stochastic 
variable obeying the (It\={o}) Langevin equation 
$\textrm{d}x/\textrm{d}t=a(x)+b(x)\eta(t)$ has probability density obeying for 
the Fokker-Planck equation 
$\partial_tP(x,t)=-\partial_x[a(x)P(x,t)]+\frac{1}{2}\partial_{xx}[b(x)^2P(x,t)]$, 
see \cite{Gardiner} for proof.

Equations~(\ref{eq:Langevin1}) and (\ref{eq:Langevin2}) underpin the main 
results of this paper, describing the evolution of a fully-connected 
spin-system growing from an initial size $N_0\gg 1$.  Their key aspects can be 
better understood by first transforming to a description in terms of the scale 
$s=N/N_0=u+v$ and magnetization $x=(u-v)/s$.  The spin flip and noise terms in 
(\ref{eq:Langevin1}) and (\ref{eq:Langevin2}) cancel to produce deterministic 
dynamics for scale,
\begin{equation}
	\frac{\mathrm{d}s}{\mathrm{d}t}=\frac{\lambda}{N_0},
	\label{eq:scale}
\end{equation}
while the magnetization remains stochastic,
\begin{eqnarray}
		\fl\frac{\mathrm{d}x}{\mathrm{d}t}=\frac{1}{s}\left(\frac{\mathrm{d}u}{\mathrm{d}t}+\frac{\mathrm{d}v}{\mathrm{d}t}\right)-\left(\frac{u+v}{s^2}\right)\frac{\mathrm{d}s}{\mathrm{d}t}\nonumber\\
		\hspace{20pt}\fl=(1-x)f_v - (1+x)f_u +\frac{\lambda}{sN_0}(g_u-g_v-x)
		+\sqrt{\frac{2(1-x)f_v+ 2(1+x)f_u}{sN_0}}\,\eta(t)\,.
	\label{eq:dm_dt}
\end{eqnarray}
We now see that growth modifies the dynamics of magnetization by \textit{i}) 
introducing a weak deterministic pull towards the magnetization of the 
reservoir, and \textit{ii}) continually reducing the magnitude of the 
stochastic fluctuations.  These apparently minor modifications can have a 
profound impact on the dynamics, as illustrated by two classic examples of 
interacting spin-systems: the voter and Glauber-Ising models.

\section{Growing voter model}
Attractive as a theoretical tool because of its lack of free-parameters, the 
surprisingly rich dynamics of the voter model arise from a simple update 
mechanism in which spins copy the state of a randomly selected neighbour.  In 
the fully-connected case, the flip probabilities are $f_u = v/s = (1 - x)/2$ 
and $f_v = u/s= (1 + x)/2$, which leads to the cancellation of the first two 
terms in (\ref{eq:dm_dt}). For non-growing systems ($\lambda=0$, $s=1$) the 
dynamical behaviour is then straightforward: the possible states are explored 
by a random walk in $x$, which halts when one of the two absorbing polarized 
states ($x=\pm1$) are reached.  This model system is non-ergodic; the time 
average of a single realisation will concentrate on just one of the polarized 
states, whereas the ensemble average will sample both with equal probability.

Things are very different when the effects of growth are included. For example, 
suppose the system is coupled to an unmagnetized reservoir ($g_u = g_v = 1/2$) 
from which spins are added at a constant rate $\lambda=1$. With no deterministic contribution, 
the equation for magnetization (\ref{eq:dm_dt}) can be decoupled from $s$ by 
introducing a rescaled time $\tau$ satisfying
\begin{equation}
	\frac{\mathrm{d}\tau}{\mathrm{d}t}=\frac{1}{s}.
	\label{eq:time_rescale}
\end{equation}
Under constant growth Eq.~(\ref{eq:scale}) gives $s=1+t/N_0$, which implies $\tau=N_0\log(1+t/N_0)$. In rescaled time (\ref{eq:dm_dt}) becomes
\begin{equation}
	\frac{\mathrm{d} x}{\mathrm{d} \tau} = -\frac{x}{N_0} + 
	\sqrt{\frac{2(1-x^2)}{N_0}}\,\eta(\tau)\,.
	\label{eq:rescaled_m_voter}
\end{equation}

\begin{figure}[t]
	\centering
	\includegraphics[height=0.15\textheight, trim=0 20 0 0]{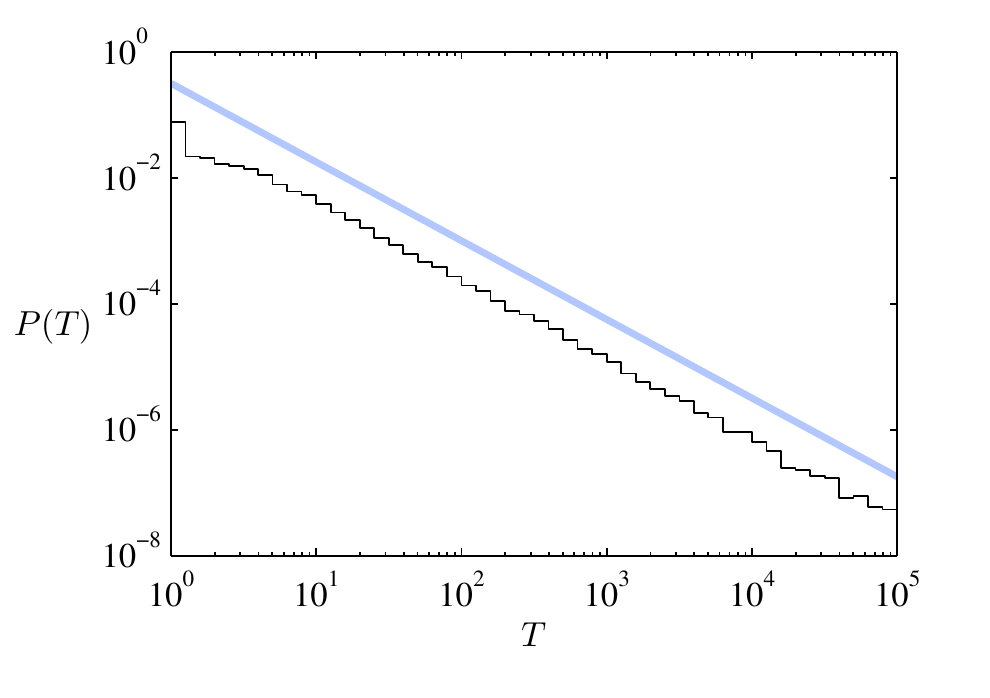}
	\caption
	{
		(Colour online) Power law tail in the distribution of first-passage time 
		$T$, for polarisation switches in a growing voter model. Black 
		histogram: simulation results from $10^4$ samples with initial size 
		$N_0=1$ ($\lambda = 1$, $g_u = g_v = 1/2$). Blue line: decay with 
		exponent -5/4 as predicted from series expansion of 
		Eq.~(\ref{eq:switch}). This result should be contrasted with the 
		non-growing case, where fully polarized states are absorbing, and 
		$T=\infty$.
	}
	\label{fig:exponent}
\end{figure}

\noindent The behaviour described by this equation is that of 
\textit{noise-induced} bistability (that is, the addition of noise changes the 
dynamics from mono- to bistable, see \textit{e.g.}, 
\cite{TBTRAJM12,TBLDAJM14}).  The noise is largest at $x=0$ when the 
deterministic pull is zero and vice-versa at the boundaries $x=\pm1$.  In the 
long-time limit, the probability distribution of the magnetization converges to 
the stationary solution of the corresponding Fokker-Planck equation. In this 
case,
\begin{equation}
\fl 0=-\partial_{x}[xP_\infty(x)]+\partial_{xx}[(1-x^2)P_\infty(x)]\quad  
\Longrightarrow\quad P_\infty(x)=\pi^{-1}(1-x^2)^{-1/2}\,.
\end{equation}
Although the distribution diverges at $x=\pm1$, these are no longer absorbing 
states, but have instead become  metastable in the sense that transitions 
between them are rare. To quantify this observation, we examine the statistics 
of the first-passage time $T$ taken to move from one polarized state to the other.  
Here, the probability distribution $f(T)$ of first-passage time obeys the 
backward Fokker-Planck equation corresponding to (\ref{eq:rescaled_m_voter}), 
but with a reflecting barrier placed at the start point and an absorbing 
barrier at the destination \cite{Gardiner}.  It was shown in \cite{TBThesis} 
that the process defined by (\ref{eq:rescaled_m_voter}) has the first-passage 
time distribution $f(\tau)=\theta_1'(0,e^{-\tau/N_0})/2\pi N_0$, where 
$\theta'_1$ denotes the derivative of the elliptic theta function of the first 
kind \cite{Elliptic} with respect to the first argument. Undoing the time 
rescaling, we find
\begin{equation}
	f(T)=\frac{1}{2\pi(N_0+T)}\theta'_1\left(0,\frac{N_0}{N_0+T}\right)\,.
	\label{eq:switch}
\end{equation}
In particular, although $T$ is almost surely finite, it has infinite mean, and 
series expansion (see \cite{Elliptic}) reveals a tail that decays as a power 
law $f(T)\sim T^{-5/4}$. This prediction is verified in continuous-time 
Monte-Carlo (Gillespie algorithm) simulations of the fully-connected voter 
model growing due to accretion of random spins, see Fig.~\ref{fig:exponent}. 

In the above example, the constant accretion of spins with random sign turned 
the voter model into an ergodic system by destroying its absorbing states.  
However, this behaviour is not shared by another natural mechanism for system 
growth: replication.  For example, suppose that the sign of newly added spins 
is determined by sampling from the existing system, so that $g_u=u/s=(1 + x)/2$ 
and $g_v=v/s=(1-x)/2$. In this case, the same time-rescaling as before can be 
applied [Eq.~(\ref{eq:time_rescale})], resulting in the autonomous equation
\begin{equation}
	\frac{\mathrm{d} x}{\mathrm{d} \tau} = 
	\sqrt{\frac{2(1-x^2)}{N_0}}\,\eta(\tau),
	\label{eq:rescaled_m_voter2}
\end{equation}
which recovers the known dynamics of a voter model of fixed size $N_0$.  
According to this equation, the system will eventually become trapped in one of 
the two absorbing states $x=\pm1$. Thus the growing system remains non-ergodic, 
in contrast to the case of growth due to the accretion of randomly aligned 
spins. If the initial magnetization is zero then both possible final states are 
equally likely, and $P_\infty(x)=\delta(x-1)/2+\delta(x+1)/2$.

\begin{figure*}[t]
	\centering
	\includegraphics[height=0.15\textheight, trim=0 10 0 0]{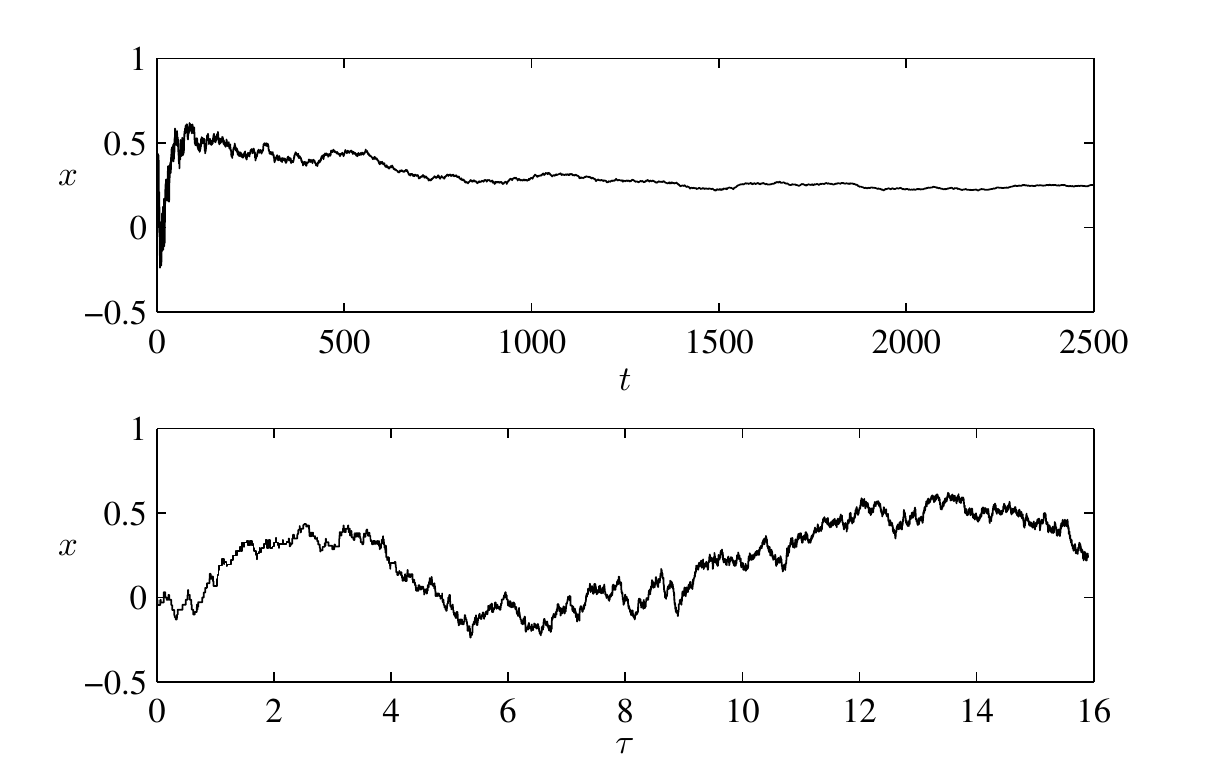}
	\includegraphics[height=0.15\textheight, trim=0 0 0 0]{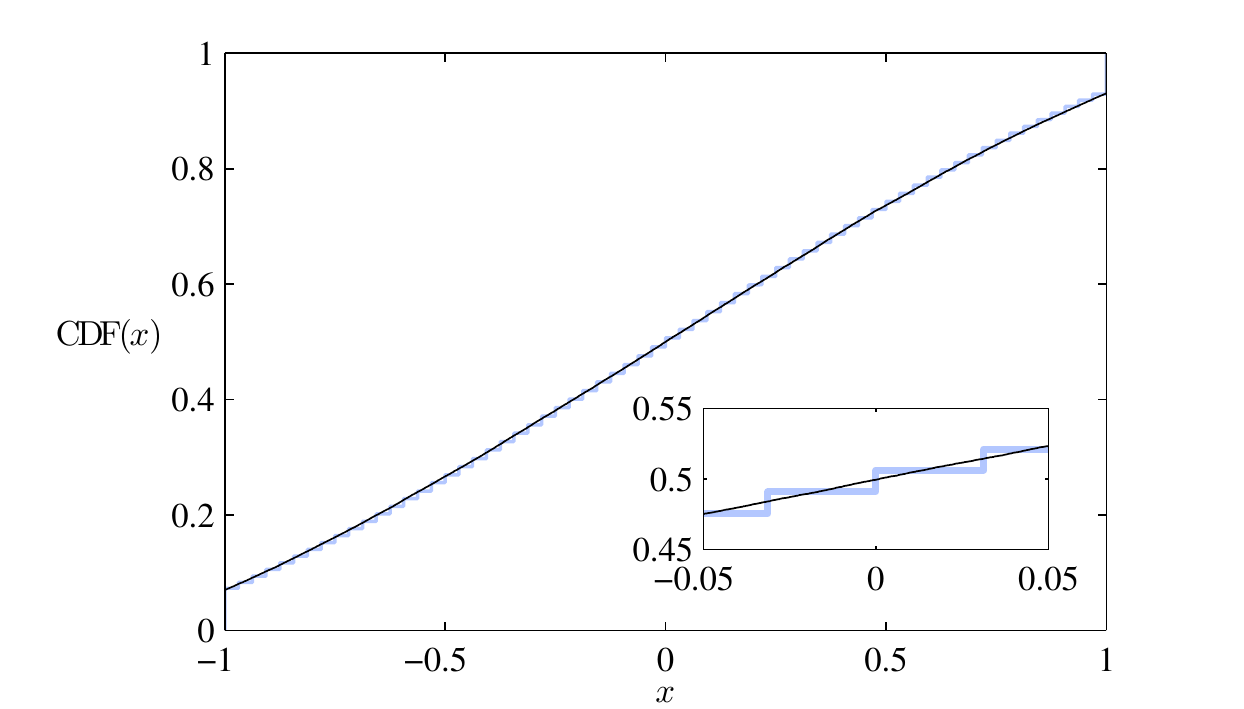}
	\caption
	{
		(Colour online) Simulation results for a growing disc 
		($\lambda=\sqrt{sN_0}$).  Left two panels: the effects of rescaling 
		time.  In growing systems, fluctuations in the magnetization $x$ due to 
		finite-size effects diminish as a function of time $t$. In rescaled 
		time $\tau$ [see Eq.~(\ref{eq:tau_2})], the system behaves like the 
		non-growing case, performing a random walk in $x$.  Right panel: freezing 
		in the Voter model at large times. The cumulative distribution function for 
		magnetization as measured from $10^4$ samples of the the growing system at long 
		times (black line) and the non-growing system at time $\tau^\ast$, in both 
		cases with $N_0=64$.
	}
	\label{fig:tau}
\end{figure*}

The treatment of growing voter models may be further generalised to cases
where $N$ is not only increasing, but \textit{accelerating}.  Within our 
framework, this is realised by a growth rate $\lambda = N^\alpha$, where 
$0<\alpha<1$.  Such situations arise, for example, in the case where growth is 
proportional to some surface area, \text{e.g.}, imagine a growing 
$d$-dimensional ball of spins, so that $\alpha = 1 - 1/d$.  (Note that 
regardless of spatial organisation the present analysis is still restricted to 
infinite range systems, \textit{i.e.}, those that are fully-connected).  If the 
sign of the incoming spins is random, the result it trivial: the deterministic 
pull towards the magnetization of the reservoir is stronger than the noise 
[since $\lambda\sim s^\alpha$ in Eq.~(\ref{eq:dm_dt})], and the system becomes 
a deterministic reflection of the reservoir in the $t\to\infty$ limit.  
However, for the case of replicating spins, the third term of (\ref{eq:dm_dt}) 
vanishes and the time-rescaling (\ref{eq:scale}) can be used once again.  Here, 
solving (\ref{eq:scale}) for $s$, substituting into the right-hand side of 
(\ref{eq:time_rescale}) and integrating, gives
\begin{equation}
	\tau = \frac{1}{\alpha}\left\{N_0^\alpha - \left[\left(1-\alpha\right)t + 
	N_0^{\alpha - 1}\right]^{\frac{\alpha}{\alpha -1}}\right\},
	\label{eq:tau_2}
\end{equation}
which has a finite limit $\tau^\ast = N_0^\alpha/\alpha$, as $t\to\infty$.  
This means that the dynamics of the accelerating voter model (with replication) 
slow asymptotically to a halt, and the probability distribution of final states 
is given by the corresponding autonomous (fixed size, rescaled-time) system, 
frozen at time $\tau=\tau^\ast$.  Indeed, the autonomous system in this case is 
precisely that described by (\ref{eq:rescaled_m_voter}), \textit{i.e.}, the 
non-growing counterpart (of size $N_0$) of the accelerating system.  
Monte-Carlo simulations presented in Fig.~\ref{fig:tau} show example results 
for a growing disc ($d=2$) with replicating surface-spins. Table~\ref{table} 
contains a summary of the four different possible growing voter model regimes.

\Table{\label{table}(Colour online) Long-time behaviours of the growing voter 
model, depending on the choice of growth rate and spin addition rule. Figures 
show the limiting distribution $P_\infty(x)$, with arrows representing Dirac 
delta functions.}
\br
 &  Accretion ($g_u=g_v=1/2$)& Replication ($g_u\hspace{-0.5pt}=u/s$, $g_v\hspace{-0.5pt}=v/s$) 
\\ \hline 
 Constant growth  & \multicolumn{1}{c}{\includegraphics[height=2cm, trim=0 0 0 
 -10]{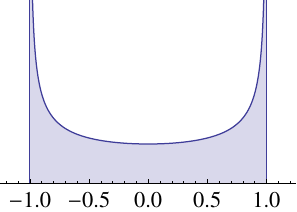}} & \multicolumn{1}{c}{\includegraphics[height=2cm, trim=0 0 0 
 -10]{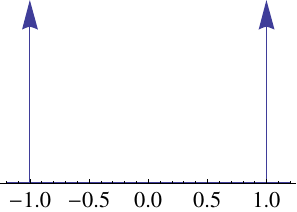}}  \\
 $(\alpha=0)$ & 
\multicolumn{1}{c}{\textit{ergodic}} & \multicolumn{1}{c}{\textit{non-ergodic}} 
\\
 \hline 
 Accelerating growth & \multicolumn{1}{c}{\includegraphics[height=2cm, trim=0 0 
 0 -10]{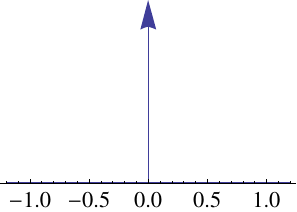}}  & \multicolumn{1}{c}{\includegraphics[height=2cm, trim=0 0 
 0 -5]{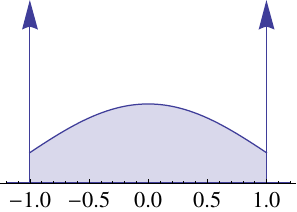}} \\ $(0<\alpha<1)$&  
 \multicolumn{1}{c}{\textit{deterministic}} & 
 \multicolumn{1}{c}{\textit{non-ergodic}} \\\br
 \end{tabular}
\end{indented}
\end{table}

\section{Growing Glauber-Ising model}
The symmetries of the voter model make it somewhat special in that its dynamics 
are driven entirely by stochastic fluctuations. What are the effects of growth 
in more general systems described by motion in a potential? To answer this 
question we analyse another canonical model of interacting spins, the 
fully-connected Glauber-Ising model \cite{Kinetic} (equivalently, the 
Curie-Weiss model with Glauber dynamics).  Here, the statistical weight 
assigned to each state is proportional to $e^{-\beta\mathcal{H}(x)}$ where 
$\beta$ is inverse temperature and $\mathcal{H}(x) = 
-\frac{J}{2}\left(Nx^2-1\right) - Nhx$ specifies the Hamiltonian, with coupling 
strength $J$ and external magnetic field $h$.  Under Glauber dynamics, the flip 
rates are given by $f_u = \frac{1}{2}\left[1 - \tanh{\beta\left(Jx + 
h\right)}\right]$ and $f_v = \frac{1}{2}\left[1 + \tanh{\beta\left(Jx + 
h\right)}\right]$ \cite{Kinetic}, such that the dynamics of finite, 
non-growing, systems are described by stochastic motion in the potential 
\begin{equation}
\Phi(x) = x^2/2 - \left(1/\beta J\right)\log\cosh\beta\left(Jx+h\right)\,,
\end{equation}
with multiplicative noise specified by state-dependent diffusion 
\begin{equation}
D(x)=N_0^{-1}\left[1-x\tanh\beta(Jx+h)\right]\,.
\end{equation}
In particular, the $N_0\to\infty$ limit of Eq.~(\ref{eq:dm_dt}) describes the 
deterministic dynamics \begin{equation}
\frac{\mathrm{d}x}{\mathrm{d}t}=\tanh\beta\left(Jx+h\right) - x\,.
\end{equation} This contrasts with the voter model, which has a flat potential 
and therefore no $O(1)$ contribution to the dynamics. The presence of a 
non-flat potential rules out decoupling $x$ and $s$ by rescaling time. 
Nevertheless, it is still possible to calculate statistical properties of 
system trajectories. 

\begin{figure}[t]
	\centering
	\includegraphics[height=0.15\textheight]{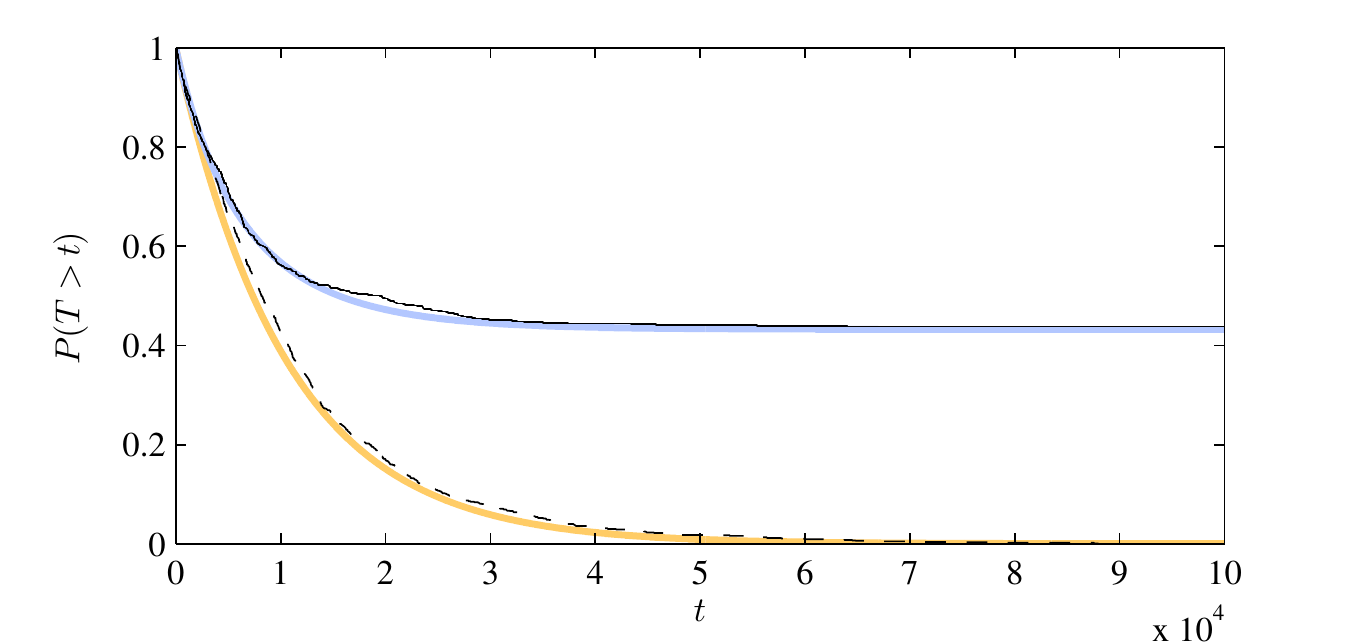}
	\caption
	{
		(Colour online) The probability $P(T\geq t)$ for a Glauber-Ising system 
		to stay stuck in the less attractive potential well up to time $t$, 
		with parameters $J=1$, $\beta=1.2$ and $h=0.002$. For a growing system 
		($\lambda=0.005$), analytical [blue line and Eq.~(\ref{eq:P_confined})] 
		and simulation (black line) results confirm that $P(T\geq t)$ converges 
		to a constant at large times. For the non-growing system [analytics 
		(yellow line) and simulation (black dashed line)] the probability of 
		remaining in the starting well decays to zero.
	}
	\label{fig:survivor}
\end{figure}

The non-growing Glauber-Ising model does not possess any absorbing states (in 
long times it samples the equilibrium distribution), however, below the 
critical temperature a pair of metastable states develop with opposite 
polarization, corresponding to minima of the potential $\Phi(x)$. This prompts 
an investigation of these states in the growing system, focusing again on the 
first-passage time $T$ to move from one potential well to the other.  Inspired 
by Kramer's method \cite{Gardiner}, a separation of timescales between two 
types of phase-space trajectory may be exploited; trajectories that travel 
between points within the same potential well are considered much more frequent 
than those which cross the potential barrier.  Under this assumption, it is 
then reasonable to introduce a size-dependent instantaneous escape rate 
$\kappa(s)$, found as the reciprocal of the \textit{mean} first-passage time 
for \emph{fixed} system size $sN_0$. This quantity can be computed via steepest 
descent.

Writing $x_0$ for the location of the starting well (without loss of generality 
we choose the left-hand well) and $x_1$ for the location of the potential 
maximum between wells, we have
\begin{equation}
x_0<x_1\,,\quad \Phi'(x_0)=\Phi'(x_1)=0\,,\quad \Phi''(x_0)>0\,,\quad 
\Phi''(x_1)<0\,.
\end{equation} We erect a reflecting barrier at $-1$ and an absorbing barrier 
at some $b>x_1$. Introducing \begin{equation}
\varphi(x)=\int_{-1}^x\frac{\Phi'(\xi)}{D(\xi)}\,d\xi\,,
\label{defvarphi}
\end{equation}
the mean first-passage time $\widetilde{T}(s)$ from $x_0$ to $b$ with fixed $s$ 
is given by \cite{Gardiner}
\begin{equation}
\widetilde{T}(s)=sN_0\int_{x_0}^b 
e^{sN_0\varphi(y)}\left(\int_{-1}^y\frac{e^{-sN_0\varphi(z)}}{D(z)}\,dz\right)\,dy\,.
\end{equation}
For large $N_0$, the inner integral here is dominated by behaviour near $x_0$ 
and the outer by behaviour near $x_1$. By Laplace's method we may approximate
\begin{equation}
\widetilde{T}(s)\approx sN_0\sqrt{\frac{2\pi}{sN_0 
|\varphi''(x_1)|}}e^{sN_0\varphi(x_1)}\left(\frac{1}{D(x_0)}\sqrt{\frac{2\pi 
}{sN_0 \varphi''(x_0)}}e^{-sN_0\varphi(x_0)}\right)\,.
\end{equation}
From the definition (\ref{defvarphi}) we compute 
$\varphi''(x)=\Phi''(x)/D(x)-\Phi'(x)D'(x)/D(x)^2$, but $x_0$ and $x_1$ are 
both turning points of $\Phi$, so the second term vanishes there. Under the 
assumption that the dynamics are approximately memoryless, the instantaneous 
escape rate is given by the reciprocal of the mean first-passage time at fixed 
size. From the above calculation we obtain \begin{equation}
\kappa(s)=\sqrt{\frac{\Phi''(x_0)|\Phi''(x_1)|D(x_0)}{4\pi^2D(x_1)}}\exp\left\{-s 
	\int_{x_0}^{x_1} \frac{\Phi'(\xi)}{D(\xi)}\mathrm{d}\xi \right\}\,.
\end{equation}
The statistics of the first-passage time $T$ of the growing system are captured 
by the \emph{survivor function}, giving the probability that the system has not 
escaped the starting well before time $t$, which is expressed in terms of 
$\kappa$ by
\begin{equation}
	P\left(T\geq t\right) = \exp\left\{-\int_0^t 
		\kappa[s(t')]\,\mathrm{d}t'\right\}\,.
	\label{eq:confined}
\end{equation}
For the case of constant growth $s=1 + \lambda t / N_0$, which, when 
substituted into (\ref{eq:confined}), gives
\begin{equation}
	P\left(T\geq t\right) = 
	\exp\left\{\frac{AN_0}{B\lambda}e^{B}\left(1-e^{B\lambda 
	t/N_0}\right)\right\}\,,
	\label{eq:P_confined}
\end{equation}
where $A=\kappa(0)$ and $B=\kappa'(0)/\kappa(0)<0$ are constants depending on 
$N_0$. In particular, we find a finite probability that the growing system may 
never leave the starting well: $P(T= \infty)>0$. This is in contrast to the 
non-growing case, which always escapes eventually.  Figure~\ref{fig:survivor} 
contains a numerical demonstration of this dichotomy.  Moreover, even if the 
growing system succeeds in switching wells once, the probability of switching 
back is even smaller (this can be seen by repeating the analysis with $N_1$ as 
the system size after the first switch). Eventually, the growing Glauber-Ising 
model is sure to become stuck in one well, thus breaking ergodicity.

\section{Discussion}
In summary, our theoretical analysis demonstrates that the interplay between 
growth and relaxation can dramatically change the behaviour of even the 
simplest of systems. We have seen how the canonical fully-connected voter and 
Glauber-Ising models respond quite differently to growth due to constant accretion of 
random spins. The fully polarized states of the voter model cease to be 
absorbing, while the metastable minima of the Glauber-Ising model eventually 
become absorbing as time goes on. A third type of behaviour was found in the 
voter model undergoing surface growth due to replication, where the system 
state converges to that of a non-growing system frozen at a finite time.  These 
results make clear the power of growth as a mechanism for driving systems away 
from equilibrium. Looking forward, one particularly interesting area of 
application lies in the field of population genetics. The fully-connected voter 
model is very closely related to the Wright-Fisher \cite{Fisher,Wright} and 
Moran \cite{Moran} models of genetic drift, in which case Table~\ref{table} 
describes the possible long-time genetic profiles of growing populations.

Finally, we have focused here on models with infinite magnetic interaction 
range.  The next step in the development of 
this new field will be the treatment of finite-range spatially heterogeneous 
systems (such as that found in recent experimental studies \cite{HHRN} of 
growing Bacterial colonies), where we can expect to uncover further novel 
phenomena driven by growth. 

\textit{Acknowledgements.} TR thanks Robert Jack and Nigel Wilding for useful 
discussion. RGM thanks George Rowlands for useful discussion, and acknowledges 
funding from EPSRC grant EP/E501311/1.

\end{document}